# Passive underwater acoustic barcodes using Rayleigh wave resonance

Yanling Zhou, Jun Fan, [a)] Jinfeng Huang, and Bin Wang

[1] *State Key Laboratory of Ocean Engineering, Collaborative Innovation Center for Advanced Ship and*

*Deep-Sea Exploration, Shanghai Jiao Tong University, Shanghai 200240, People's Republic of China*

**Abstract:** A passive underwater acoustic marker is presented and its feasibility for underwater recognition, positioning, and navigation is proved by a numerical method and experimental results. These markers are composed of acrylic elastic objects designed by backscattering strong peaks associated with the subsonic Rayleigh wave resonance of a polymer target excited by a broadband pulse, and having a unique acoustic signature for a selected frequency band, akin to acoustic barcodes. Therefore, the backscattering response of markers can be regulated by changing the geometry of elastic objects. These acoustic barcodes naturally operate in a wider frequency band, and have a longer lifetime and lower cost, compared with active acoustic markers.

---

[a)] Author to whom correspondence should be addressed: fanjun@sjtu.edu.cn

# 1. Introduction

Underwater acoustic communication technology is the most effective in water, and acoustic fiduciary markers have been applied for recognition, positioning, and navigation in various fields of offshore operation in addition to the black box on aircraft and ships. It is particularly important to aid in navigating autonomous underwater vehicles, and localization of undersea mines, multi-target cooperation, and the salvage of crashed planes and sunken ships using underwater acoustic beacons.[1] Typically, active acoustic beacons acting as a primary navigation method continuously emit acoustic pulse signals, which can be expensive to maintain and have extremely short operating times. However, the transducer, power amplifier, and electronic cabin increase the load of targets. Therefore, a passive underwater acoustic marker that responds only when interrogated is needed, which is inexpensive in operation and deployment, and can handle varying environmental conditions. Passive acoustic identification (AID) markers offer a simple and effective solution to these design requirements. Similar to the function of optical barcodes, AID markers are designed to operate as acoustic barcodes. In recent years, researchers have explored a host of approaches for acoustic coding techniques, most of which focus on air acoustics.[2–4] In underwater acoustics, Srivastava et al.[5] demonstrated the Bragg scattering characteristics of periodically engraved surfaces on generating a unique three-dimensional acoustic signature with designed spatial and frequency variations that can be detected by a high-frequency SONAR. Thus, it is challenging to sweep over the marker in the space domain. Satish et al.[6] introduced an alternative passive marker made of horizontally stratified layers of different elastic or viscoelastic materials with varying bulk acoustic properties and thicknesses, wherein a unique acoustic signature was generated specific to the selection of the materials and thickness of each layer for the desired frequency pulse. Hence, akin to an acoustic barcode, the marker signature may be uniquely engineered by varying the thickness and acoustic properties of the layers. Since the designed marker is planar, any off-normal incident wave cannot be detected by a monostatic source–receiver configuration. To overcome this effect, a modified passive marker designed with radially stratified spherical shells is proposed, the specular backscatter of which is independent of the incidence orientation.[7] These approaches rely on markers' unique structural responses (i.e., geometric scattering characteristics); however, they are limited to high-frequency SONAR systems, and their acoustic wave attenuation is much faster, which is a challenge for remote detection, recognition, and positioning.

Inspired by SonarBell,[8] many studies on the acoustic scattering of polymer targets have been carried out. It is shown that the backscattering response exhibits strong resonance peaks that arise from the resonance of subsonic Rayleigh waves in the frequency domain. Based on this characteristic, a passive acoustic marker (also referenced as "acoustic barcode" in this study), composed of double acrylic spheres, is proposed, which has a unique acoustic signature for the selected frequency regime. A specific elastic structure is designed to generate single or multiple resonance peaks in the bandwidth *B* of the echo divided into *m* uniform subbands, and these peaks are distributed in different subbands, as shown in Fig. 1. The bar is black when the resonance peak exists in the subband, and is otherwise expressed as white, thus generating a black and white acoustic barcode. Each "1" is represented as the resonance peak that existed in the subband; conversely, each "0" is a blank (i.e., no resonance peak). Results of numerical calculation and experiment validate the feasibility of the proposed concept of an acoustic barcode composed of double acrylic spheres. Each marker generates a unique acoustic scattering signature that arises from the subsonic Rayleigh wave resonance that can be identified using a stored library of pre-computed or experimentally calibrated responses. This allows more time for maritime rescue and salvage in case of placing a passive marker on aircraft and ships. Furthermore, analogous to the traditional barcode, the information is modulated in the acoustic scattering characteristics of the target, which achieves covert information transmission and target identification.

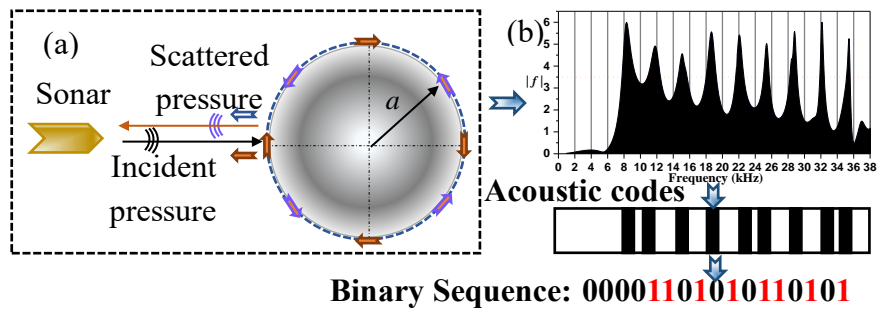

Fig. 1. (Color online) (a) Schematic of propagation physics involved in backscattering response of a sphere. Incident wave generates a specular scattering wave (open arrows)

and an elastic scattering wave (solid arrows). (b) Amplitude–frequency characteristics of backscattering form function for incident wideband pulse. Black bars correspond to the resonance peak that existed in the subband, while white bars represent no resonance peak. Black and white bars are processed, resulting in a binary sequence.

## 2. Numerical simulations

Based on the approximate ray synthesis, Hefner and Marston[9] explained the physical mechanism of a large backscattering enhancement for a solid acrylic sphere that is associated with the subsonic Rayleigh wave, and found that the resonance peaks of the backscattering form function are strong at low frequency. In view of the Rayleigh wave phase-matching method, the approximate formula of resonance frequencies for an acrylic sphere is given by[10]

$$f_0 = \frac{2n+1}{2} \frac{c_R}{2\pi a}, \tag{1}$$

where $n=0, 1, 2, 3…$, $c_R$ is the Rayleigh wave speed, and $a$ the radius.

It can be seen that the resonance frequencies of the backscattering form function are dependent on the Rayleigh wave speed and the radius of the sphere according to Eq. (1). The phase velocity of the Rayleigh wave increases with increasing frequency and finally tends to a constant when the effects of Rayleigh wave dispersion can be ignored. The Rayleigh wave velocity is a determined value for a desired frequency range, and thus the backscattering resonance peaks can be regulated by varying the radius of spheres or the combination of spheres of different radius. The acoustic scattering signature is unique for each combination, so the acoustic coding can be realized over the entire frequency range of interest. A passive acoustic marker composed of double spheres [see Fig. 2(a)] is proposed, the centers of which are located on the *z* axis. In this model, the target is axi-symmetric, while the incident acoustic field is plane wave and non-axi-symmetric. To reduce the amount of calculation, these numerical simulations can be implemented by using the axi-symmetric acoustic scattering calculation method by incorporating it into COMSOL Multiphysics® (Comsol, Inc., USA) field coupling software.

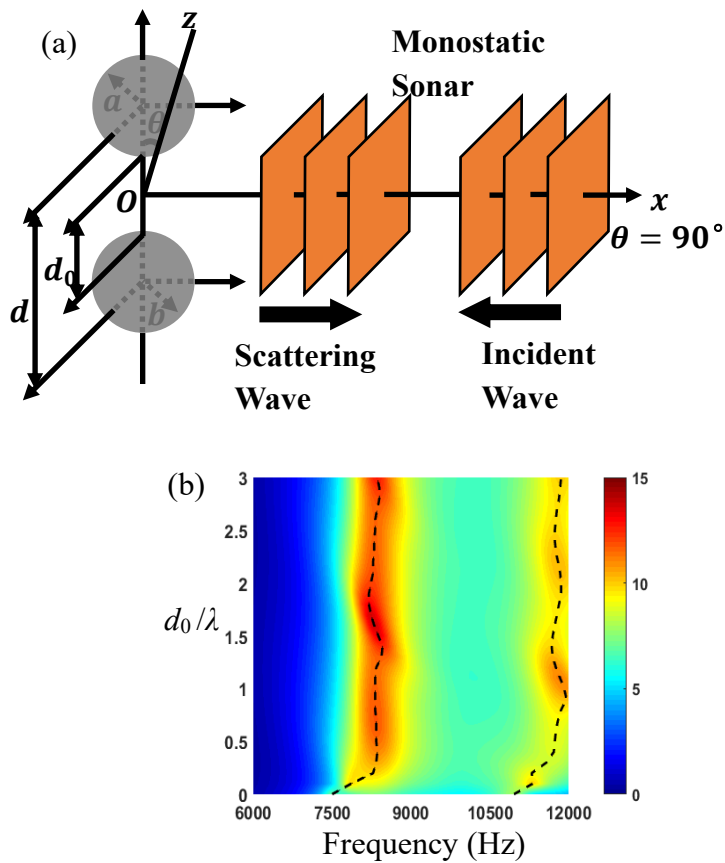

Fig. 2. (Color online) (a) Scattering diagram of two spheres at plane normal incidence (i.e., $\theta$=90°) in monostatic configuration; radii of the two spheres are *a* = 0.05 m. The density of the sphere is $\rho$ = 1190 kg/m³, and the computed values of compressional velocity and shear velocity are $c_L$ = 2690 m/s and $c_T$ =1340 m/s, respectively. Distance between nearest vertices of double spheres is $d_0$; therefore, distance between centers of the two spheres is $d = d_0 + 2a$. (b) Frequency–distance spectrum of backscattering form function for two spheres using finite-element method. Black dashed line represents resonance frequencies of backscattering form function varying the ratio of the distance between two spheres and the incident wavelength $d_0/\lambda$.

The plane acoustic wave is incident perpendicular to the direction of the connecting line between the centers of two spheres (i.e., $\theta$ = 90°) in the monostatic configuration, and the frequency-distance spectrum of the backscattering form function for two spheres is shown in Fig. 2(b). The color represents the amplitude of the form function, horizontal coordinate is the frequency (Hz), and vertical coordinate is the ratio of the distance between two spheres and the incident wavelength (i.e., $d_0 / \lambda$) ranging from 0 to 3 in intervals of 0.1. The black dashed line represents the resonance frequencies varying the ratio of the distance between two spheres and the incident wavelength. It can be seen that the first resonance frequency increases with distance when $d_0 < \lambda/2$, and changes slightly in the range 7500–9000 Hz in the case of $\lambda/2 \ll d_0 \leq 2\lambda$, and remains constant with increasing distance when $d_0 \gg 2\lambda$; that is, the interaction between the two spheres (occlusion, multiple scattering, and others) can be ignored. Therefore, the distance is $d_0$ = 0.25 m using the combination of the two spheres for acoustic coding.

The frequency spectrum of the broadband signal over the range 6–10.5 kHz is divided into three uniform subbands, and each subband width is $\Delta f = 1.5$ kHz. Based on the first resonant peak of the backscattering form function for an acrylic sphere (i.e., $ka \approx 1.74$), spheres with radii of $a_1 = 0.045$ m, $a_2 = 0.05$ m, $a_3 = 0.06$ m, and $a_4 = 0.065$ m are separately selected. After the combination of two spheres with a radius selected as above, seven combinations and acoustic coding results are obtained, as shown in Table 1.

Table 1. Combinations of spheres of different radii.

| Combinations | $a_1$ - $a_1$ | $a_2$ - $a_2$ | $a_1$ - $a_4$ | $a_1$ - $a_3$ | $a_2$ - $a_4$ | $a_1$ - $a_2$ | $a_2$ - $a_3$ |
|---|---|---|---|---|---|---|---|
| Resonance frequency (kHz) | 9.2 | 8.24 | 6.38 | 6.92/9.44 | 6.38/8.6 | 8.28/9.32 | 6.92/8.5/9.44 |
| Binary codes | 001 | 010 | 100 | 101 | 110 | 011 | 111 |

According to formula (1), the resonance frequencies of the selected combinations can be quickly estimated. Thus, the backscattering form function amplitude of each combination is obtained as shown in Fig. 3. A resonance peak (red pentagram in Fig. 3) is found and the amplitude is larger than the threshold (i.e., $|f| = 6$ (dotted line in Fig. 3) determined by the amplitude corresponding to the maximum decrease of 6 dB) in the subband, when the barcode is black; otherwise, it is white, which forms a black and white acoustic barcode. Then, applying existing binarization approaches to the barcode, the acoustic coding with a combination of "0" and "1" over this bandwidth can be obtained in the frequency domain. Analogous to traditional barcodes, this kind of code generated in the frequency domain can be called an underwater acoustic barcode. The elastic structure designed to generate acoustic barcodes is known as an acoustic barcode generator.

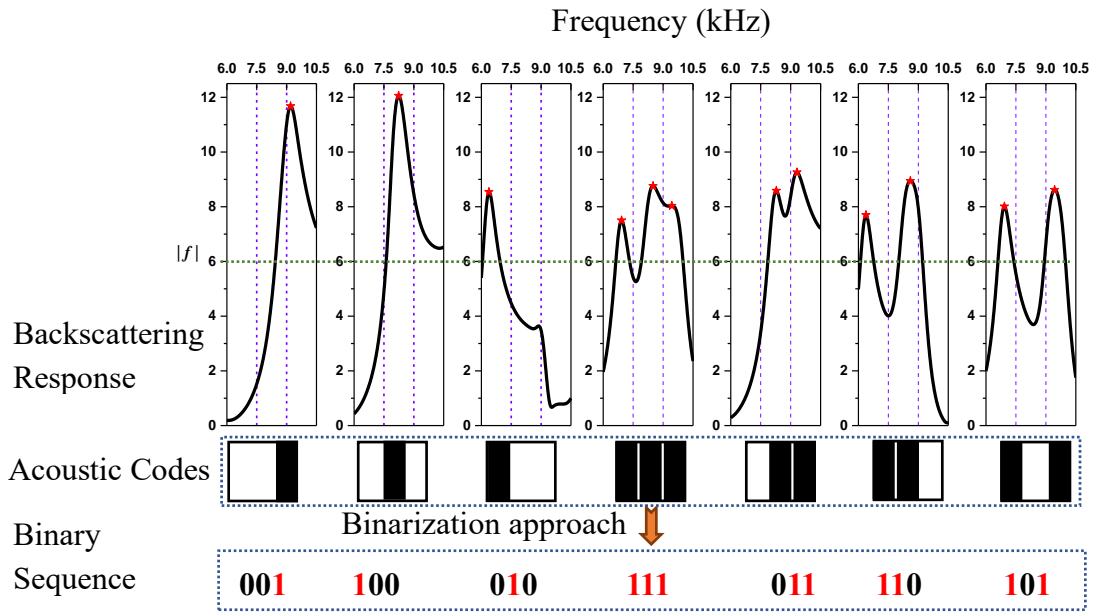

Fig. 3. (Color online) Acoustic barcodes combined by two spheres. Upper panel is backscattering form function of the two spheres in various combinations listed in Table 1, middle panel shows acoustic barcodes, and lower panel shows the binary sequence compiled by barcodes.

## 3. Experiments

To validate the proposed concept for the design of the passive underwater acoustic marker composed of two acrylic spheres, a water tank measurement of acoustic scattering was conducted. According to the combinations of sphere radius in Table 1, the centers of each two spheres are connected by nylon screw rods and seven experimental objects are formed [see Fig. 4(a)]. A linear frequency-modulated pulse with a frequency range of 6–10.5 kHz and pulse width of 0.5 ms was projected onto the experimental object from the transducer, and the pulse period was 500 ms. The transducer was maintained at distances of 3.65 and 1.3 m from the test specimen and receiver (B&K 4038), respectively. The geometric centers of the transducer, test object, and hydrophone were immersed at the same depth, i.e., 4 m, in a 30 m × 20 m × 10 m water-filled tank; the experimental setup is shown in Fig. 4(b). In the monostatic configuration, the target rotated at a constant, computer-controlled speed of 45°/min. The entire measurement system was powered by a high-capacity uninterruptible power supply to eliminate any power-line interference. The backscattering response of the combined spheres was obtained at different incident azimuths. The form function of the two spheres in various combinations was then extracted at the angle $\theta = 90°$, as expected, and showed excellent agreement with numerical results. According to the method outlined in section 2, the binary information is encoded by assigning different bits to each combination related to resonance peak; that is, there is a resonance peak in the subband, the value of which can be specified as "1", otherwise it is "0". The recognition and positioning of specific underwater targets can be realized by mobilizing the pre-calculated or experimental calibration response repository.

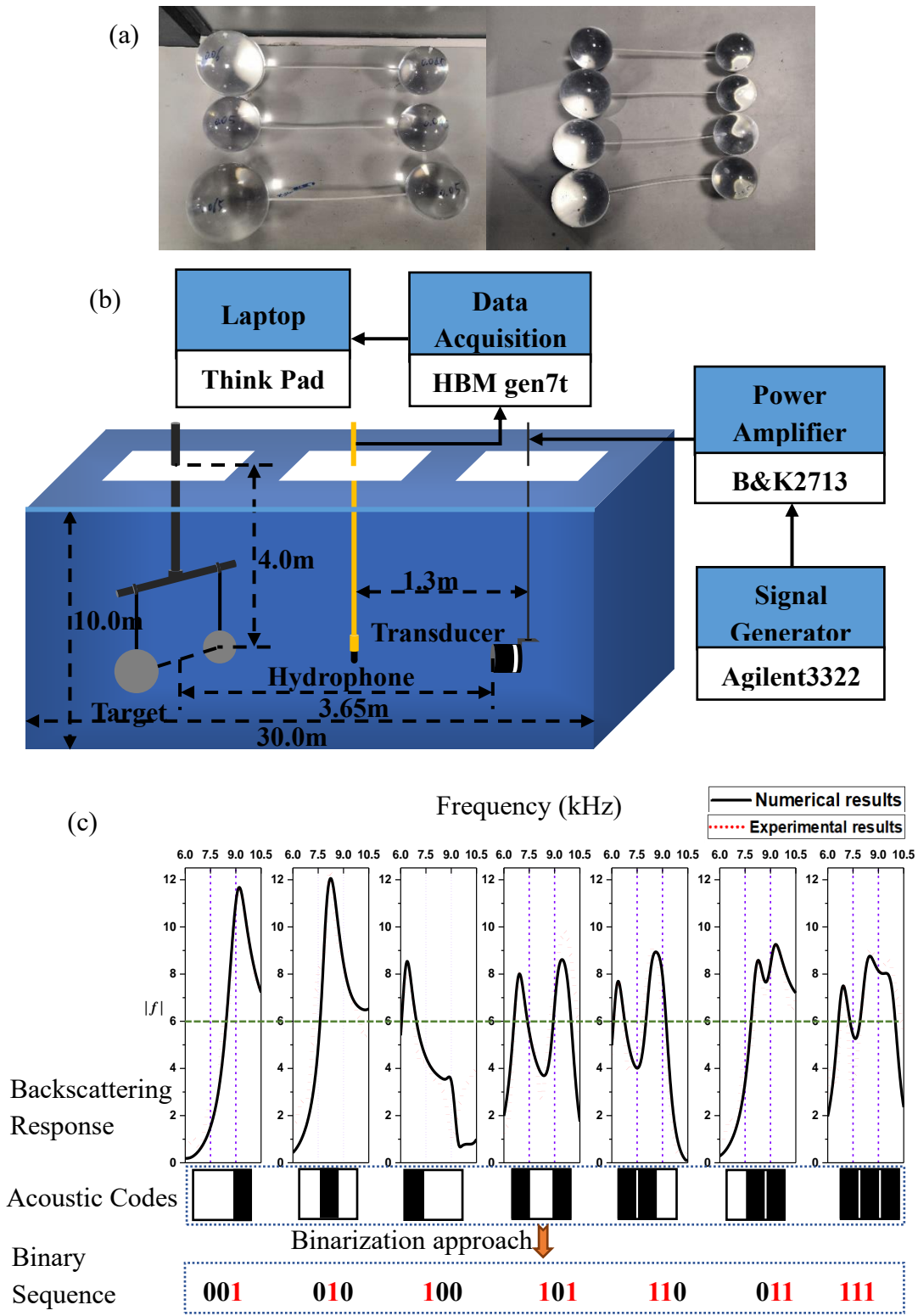

Fig. 4. (Color online) (a) Photograph of experimental objects. Radii of seven combinations of spheres are listed in Table 1. (b) Sketch of experimental setup. (c) Comparison of numerical and experimental results of backscattering response in various combinations.

## 4. Results and discussion

The time-domain signal for an incident angle of $\theta = 90°$ corresponding to the backscattering response was extracted. Using broadband signal processing, the backscattering form functions of the experimental model in various combinations in the frequency range 6–10.5 kHz were obtained, as shown in Fig. 4(c). Owing to the processing accuracy of the experimental model, material parameter error, and test error, some errors exist between the numerical calculations and experimental results. However, the results of numerical calculation [black solid line in Fig. 4(c)] and experiment (dotted line in same figure) are in agreement well within the allowable range of error, especially the resonance peak. This demonstrates the effectiveness of acoustic coding using the frequency-domain resonance characteristics of combined spherical targets. Additionally, the backscattering enhancement and smaller attenuation of the acoustic wave at low frequencies endow the acoustic marker of the combined acrylic spheres with the advantage of long-distance detection, identification, positioning, and navigation. More generally, the proposed method can be used to specify acoustic markers for any frequency range of interest as long as the radius of each sphere is appropriately selected.

## 5. Conclusions

The acoustic barcodes devised by combining two acrylic spheres presented in this paper constitute a proof of concept for the design of passive underwater acoustic markers that can produce a unique acoustic scattering signature arising from the subsonic Rayleigh wave resonance. The backscattering response from these markers is determined by the radius and material characteristics of the combined-sphere markers. Furthermore, it was demonstrated that the acoustic signatures of these multisphere-based acoustic markers can be predicted rapidly using a simple approximate formula that allows one to select optimal combinations of material parameters and the radius of each sphere in composing a tag to ensure the uniqueness of the acoustic signatures for a given frequency band, akin to a traditional bar code, according to the intended purpose of acoustic markers (e.g., navigation or information coding). Subsequently, experiments on the acoustic scattering of two spheres were conducted to experimentally validate the proposed concept by comparing the measured responses with simulation results. In addition, the geometry based on the resonance characteristics of elastic waves is used for passive acoustic marking, including, but not limited to, a composite cylinder, cylinder shell, or spherical shell, which can also be studied in a similar way. Therefore, through the fine design of the elastic structure, the resonance peaks of the scattering response within a certain bandwidth can be adjusted and then used for acoustic coding. The underwater acoustic barcode generator can be installed on the underwater target, such as an autonomous underwater vehicle, underwater aquaculture cage, and others, which can be used as the identification mark of an underwater target.

## Acknowledgments

This work was supported by the National Natural Science Foundation of China Grant (Grant No. 11774229). The experimental data was obtained under the guidance of Prof. Jun Fan, and also with the help of laboratory technician Kai Xu, Yefeng Tang, Wentao Wang, Shaobo Wang. Finally, the authors wish to acknowledge the anonymous reviewers for the insightful comments and helpful suggestions made.

## References and links